\documentclass[twocolumn]{aastex631}

\usepackage{graphicx}	
\usepackage{amsmath}	
\usepackage{amssymb}	
\usepackage{comment}   
\usepackage{booktabs}  %
\usepackage{hyperref}
\usepackage{float}
\usepackage{quoting}
\usepackage{longtable}
\usepackage{threeparttable}
\usepackage{threeparttablex}
\usepackage{xcolor}
\usepackage{ragged2e}

\begin{document}

\title{Exploring the archives: A search for novae in UVIT snapshots of M31}

\correspondingauthor{Judhajeet Basu}
\email{judhajeet20@gmail.com}

\author[0000-0001-7570-545X]{Judhajeet Basu}
\affiliation{Indian Institute of Astrophysics, 2nd Block Koramangala, 560034, Bangalore, India}
\affiliation{Pondicherry University, R.V. Nagar, Kalapet, 605014 Puducherry, India}

\author[0009-0009-8534-9765]{Krishnendu S.}
\affiliation{Indian Institute of Astrophysics, 2nd Block Koramangala, 560034, Bangalore, India}
\affiliation{Amrita School of Physical Sciences, Amrita School of Engineering, Amrita University, 641112, Ettimadai, India}

\author[0000-0002-3927-5402]{Sudhanshu Barway}
\affiliation{Indian Institute of Astrophysics, 2nd Block Koramangala, 560034, Bangalore, India}

\author[0009-0000-5909-293X]{Shatakshi Chamoli}
\affiliation{Indian Institute of Astrophysics, 2nd Block Koramangala, 560034, Bangalore, India}
\affiliation{Pondicherry University, R.V. Nagar, Kalapet, 605014 Puducherry, India}

\author[0000-0003-3533-7183]{G.C. Anupama}
\affiliation{Indian Institute of Astrophysics, 2nd Block Koramangala, 560034, Bangalore, India}

\begin{abstract}
Extensive multi-wavelength studies of novae have been carried out in our galaxy and in M31 for decades. However, UV studies of extragalactic novae are limited, especially those in quiescence. For the first time, we present a UV catalog of novae in M31 using the archival AstroSat UVIT imaging data. We used two image subtraction techniques to retrieve objects located deep into the M31 central region. We have found 42 novae in total in the UVIT images, 15 of which have been detected in multiple filters in FUV and NUV. The novae detected at quiescence show signatures of accretion disk from their UV spectral energy distributions, whereas those in the outburst phase show signatures of pseudo-photosphere. A few novae were also detected in multiple epochs. Some show a near-constant FUV magnitude at quiescence, while others caught near the outburst reveal pre-eruption dips in their light curves. We conclude with a discussion on the significance of UV surveys in illuminating theoretical predictions for novae systems, including detecting the elusive early UV flash.

\end{abstract}

\keywords{galaxies: individual (M31) -- novae, cataclysmic variables -- stars: techniques: image processing -- techniques: photometric -- transients: novae}

\section{Introduction} \label{sec:intro}

Novae are cataclysmic events triggered by thermonuclear runaway reactions on the surface of accreting White Dwarfs (WD) \citep{Starrfield_1999, Bode_2008, Starrfield_2016}. These novae eruptions increase the brightness of the binary system up to several orders of magnitude, triggering nucleosynthesis and expelling material into the interstellar medium (ISM) \citep{Hernanz_1998, Gehrz_98}. Individual novae are excellent laboratories for understanding accretion and mixing processes \citep{Guo_2022}, binary evolution \citep{Kraft_1964, shen_2022}, supernova type~Ia progenitors \citep{Hill16, wang18}, and shock mechanisms \citep{aydi_2020,chomiuk_2021}. Nova studies are also important tracers of galactic chemical evolution and ISM enrichment \citep{Starrfield_2020}.

One of the galaxies in which novae are studied in detail is M31, owing to full visibility for more than half a year, close distance, low inclination, and high nova rate. High occurrences of recurrent novae in M31, particularly those with short recurrence periods of fewer than 10 years such as M31N2008-12a \footnote{All M31 novae denoted by M31NYYYY-MMx will be denoted by YYYY-MMx hereafter} \citep{basu_2024a}, 2017-01e, 2013-10c, 1990-10a, 1963-09c and others have intrigued astronomers to explore the M31 galaxy. M31 has been surveyed extensively for novae for several decades in optical wavebands \citep{Arp_1956, Rosino_1964, Rosino_1973, Ciardullo_1987, Rosino_1989, Sharov_1991, Tomaney_1992, Rector_1999, Shafter_2001, darnley_2004, Darnley_2006, Kasliwal_2011, Shafter_2011a, Lee_2012, Williams_2014, Williams_2016, Rector_2022}. Over the last two decades, a few extensive studies have been performed in X-rays as well \citep{Peitsch_2007, Pietsch_2010b, Hen10, Hen11, Hen14}. However, surveys are limited in IR \citep{Shafter_2011} and UV \citep{cao_2012} bands. Quiescence studies of M31 novae were conducted by \citet{Williams_2014, Williams_2016} to look for progenitors using HST optical data, taking advantage of its high resolution. 

Galactic novae have been studied in the UV since the launch of IUE (\citealt{Starrfield_1986} and references therein). \citet{Selvelli_2013} studied a sample of 18 galactic novae in quiescence in UV, focusing on their spectral energy distributions (SEDs), which were later combined with outburst phase data by \citet{Selvelli_2019} for an overall understanding of the properties of old novae. \cite{page_2020,Page_2022} summarized the UV and X-ray light curves of galactic novae observed by \textit{Swift}. 

Rare UV studies of extra-galactic novae, such as \citet{cao_2012} for M31, and \citet{Lessing_2023} and \citet{Shara_2023} for M87, concentrated primarily on eruption characteristics, spatial distribution, nova rates, and population studies. In the case of the Magellanic Clouds, their wide fields and low nova rates \citep{Mroz_2016} have led to limited surveys in optical and other bands. However, focusing on some individual novae, for instance, SMCN 2016-10a \citep{Aydi_2018}, Nova LMC 1971b (\citealt{Bode_2016}), and Nova LMC 1968 (\citealt{Kuin_2020}) have revealed their UV quiescence features.  Nonetheless, no study has been reported in the literature honing on the systematic study of the quiescence phase of extragalactic novae in UV. Though nova systems in M31 and other galaxies are expected to be similar to those in our Milky Way, the missing gap must be filled for completeness. 

This work presents the search for novae in M31 in the UV bands using archival AstroSat-UVIT data. 
The paper discusses our data reduction, analysis, and detection methods and provides a catalog of the novae detected (including those caught during outbursts) as well as non-detection upper limits. We also briefly discuss the importance of UV missions and multi-wavelength facilities as synoptic surveys for novae studies.

\section{Data} \label{sec:UVIT data red}

\subsection{Observations and data reduction}

\begin{table*}
    \caption{UVIT observation log of all images, including co-added frames, in which at least one nova was detected. The full table, which includes details of all individual and co-added frames analyzed in this work, will be available in the published version.}
    \label{obs_table}
    \centering

    \begin{tabular}{ccccccccc}
    \toprule
        Observation ID & RA & DEC & Filter & Field & Obs Epoch & Exp time & Limiting Mag & Novae detected \\  
                     & (deg) & (deg) &     &       &   (BJD) & (s)   &    (AB)       &                \\
        \midrule
        A02\_028T01\_9000000724 & 10.71683 & 41.12430 & F148W & M31 01 & 2457671.85662 & 7736.347 & 22.94 & 18 \\ 
        A02\_028T01\_9000000724 & 10.71683 & 41.12430 & F172M & M31 01 & 2457672.19496 & 3611.883 & 21.39 & 14 \\
        A02\_028T01\_9000000724 & 10.71683 & 41.12430 & N219M & M31 01 & 2457671.85657 & 7780.686 & 22.08 & 12 \\
        A02\_028T01\_9000000724 & 10.71683 & 41.12430 & N279N & M31 01 & 2457672.19491 & 3627.438 & 21.53 & 11 \\  
        A07\_007T04\_9000003310 & 10.70412 & 41.27929 & F148W & M31 01 & 2458805.25645 & 17045.731 & 23.37 & 17 \\  
        A07\_007T04\_9000003310 & 10.70412 & 41.27929 & F169M & M31 01 & 2458804.85075 & 10426.754 & 22.63 & 15 \\  
        A07\_007T04\_9000003310 & 10.70412 & 41.27929 & F172M & M31 01 & 2458806.27133 & 16606.192 & 22.34 & 15 \\  
        A10\_002T04\_9000004022 & 10.59843 & 41.21609 & F148W & M31 01 & 2459174.71396 & 11759.864 & 23.10 & 18 \\  
        A02\_028T03\_9000000788 & 11.01776 & 41.54606 & F148W & M31 02 & 2457704.13548 & 7942.829 & 22.69 & 3 \\  
        A02\_028T03\_9000000788 & 11.01776 & 41.54606 & F172M & M31 02 & 2457704.48574 & 5350.934 & 21.48 & 2 \\  
        A02\_028T03\_9000000788 & 11.01776 & 41.54606 & N219M & M31 02 & 2457704.13543 & 7986.231 & 22.01 & 2 \\  
        A02\_028T03\_9000000788 & 11.01776 & 41.54606 & N279N & M31 02 & 2457704.48567 & 5544.727 & 21.33 & 3 \\  
        A10\_002T05\_9000004000 & 11.13728 & 41.48918 & F148W & M31 02 & 2459167.97966 & 12610.195 & 22.77 & 1 \\  
        A04\_022T02\_9000002184 & 11.33933 & 41.88125 & F148W & M31 08 & 2458292.30948 & 3371.811 & 22.85 & 1 \\  
        A04\_022T02\_9000002184 & 11.33933 & 41.88125 & F172M & M31 08 & 2458292.98608 & 9017.513 & 22.29 & 1 \\    
        A05\_004T03\_9000002508 &  9.87699 & 40.36606 & F148W & M31 15 & 2458435.07551 & 4610.422 & 23.04 & 1 \\  
        A05\_004T03\_9000002508 &  9.87699 & 40.36606 & F169M & M31 15 & 2458435.27774 & 9286.96 & 22.69 & 1 \\  
        A05\_004T03\_9000002538 &  9.86651 & 40.37805 & F148W & M31 15 & 2458449.21674 & 2104.392 & 22.72 & 1 \\  
        A05\_004T03\_9000002538 &  9.86651 & 40.37805 & F169M & M31 15 & 2458449.42980 & 9224.752 & 22.76 & 1 \\
        \midrule
        Co-added frames $^*$    & 10.70959 $^\dag$ & 41.27957 $^\dag$ & F148W & M31 01 & 2458550.60901 $^\ddag$ & 36541.942 & 23.54 & 3 $^\#$ \\
        Co-added frames $^*$    & 10.70959 $^\dag$ & 41.27957 $^\dag$ & F172M & M31 01 & 2458239.23315 $^\ddag$ & 20218.075 & 22.50 & 1 $^\#$\\
        \bottomrule
    \end{tabular}

    \raggedright
    \textbf{Notes} \\
    $^*$ Co-added frames include all the images of the same field in the same filter combined together. For example, the co-added frame of the M31 01 field in the F148W filter includes the sum of F148W images with Obs ID A02\_028T01\_9000000724, A07\_007T04\_9000003310 and A10\_002T04\_9000004022. \\
    $^\dag$ RA DEC of the coadded frames are the centers of the co-added images. \\
    $^\ddag$ Observation epochs for co-added frames are the mid-point of the first and last observation. \\
    $^\#$ The number of novae detected in the co-added frames but not in the individual images. 

\end{table*}

\textit{AstroSat} is a space-based high-energy telescope in a low inclination Earth orbit \citep{KPS14}. The Ultraviolet Imaging Telescope (UVIT) is one of the primary instruments onboard \textit{AstroSat}, consisting of a twin telescope with one FUV and one NUV/VIS channel, capable of observing simultaneously. Both telescopes have a mirror of 38 cm diameter and a circular field of view of 28 arcmin. UVIT has one of the best spatial resolutions at 1.5 arcsec in UV, making it suitable for studying extragalactic UV sources. UVIT's FUV filters offer a unique waveband (up to 1000 \AA) to detect and study novae. Ground-based and in-orbit calibrations of the UVIT instrument are reported in \citet{annapurni16} and \citet{tandon17a,tandon17b}. 

UVIT has observed multiple fields of M31 in different filters over multiple epochs. The observations and datasets summarized by \cite{leahy_2020, Leahy_2021} are used in this work. Additional fields observed later by the same PI and other M31 fields observed by other PIs have also been looked into for this work. All the data utilized in this work have been summarised in Table~\ref{obs_table}. The \texttt{Level 1} and \texttt{Level 2} data have been made available online by the Indian Space Science Data Center's (ISSDC) \textit{astrobrowse}\footnote{\href{https://astrobrowse.issdc.gov.in/astro_archive/archive/Home.jsp}{https://astrobrowse.issdc.gov.in/astro\_archive/archive/Home.jsp}}. We obtained all the \texttt{L1} data and processed it using \texttt{CCDLAB} \citep{postma17}. The automated steps in \texttt{CCDLAB} include flat-fielding, drift correction, and cosmic-ray correction, among others discussed in detail in \citet{postma21}. The orbit-wise images were registered and merged to get a single image with a high signal-to-noise ratio. The final step in \texttt{CCDLAB} involved performing astrometry on the merged and orbit-wise images.

\begin{figure*}[t]
    \centering
    \includegraphics[scale=0.33]{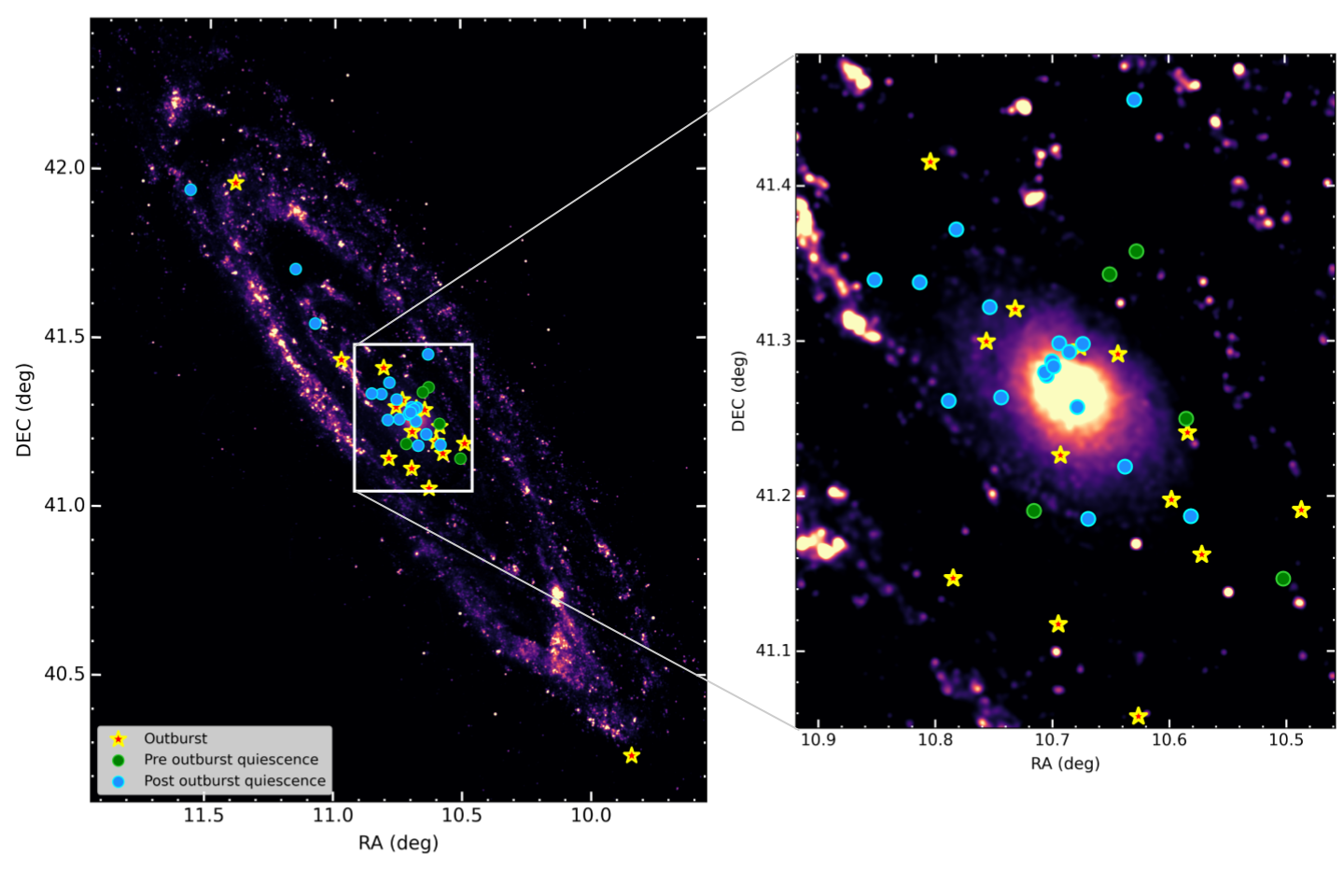}
    \caption{UVIT F148W mosaic image of M31 generated using \texttt{SWarp}. The locations of all the novae detected in the archival images are marked in the figure. The inset shows a zoomed-in version of the M31 central region.}
    \label{fig:M31_with_novae}

    \includegraphics[scale=0.25]{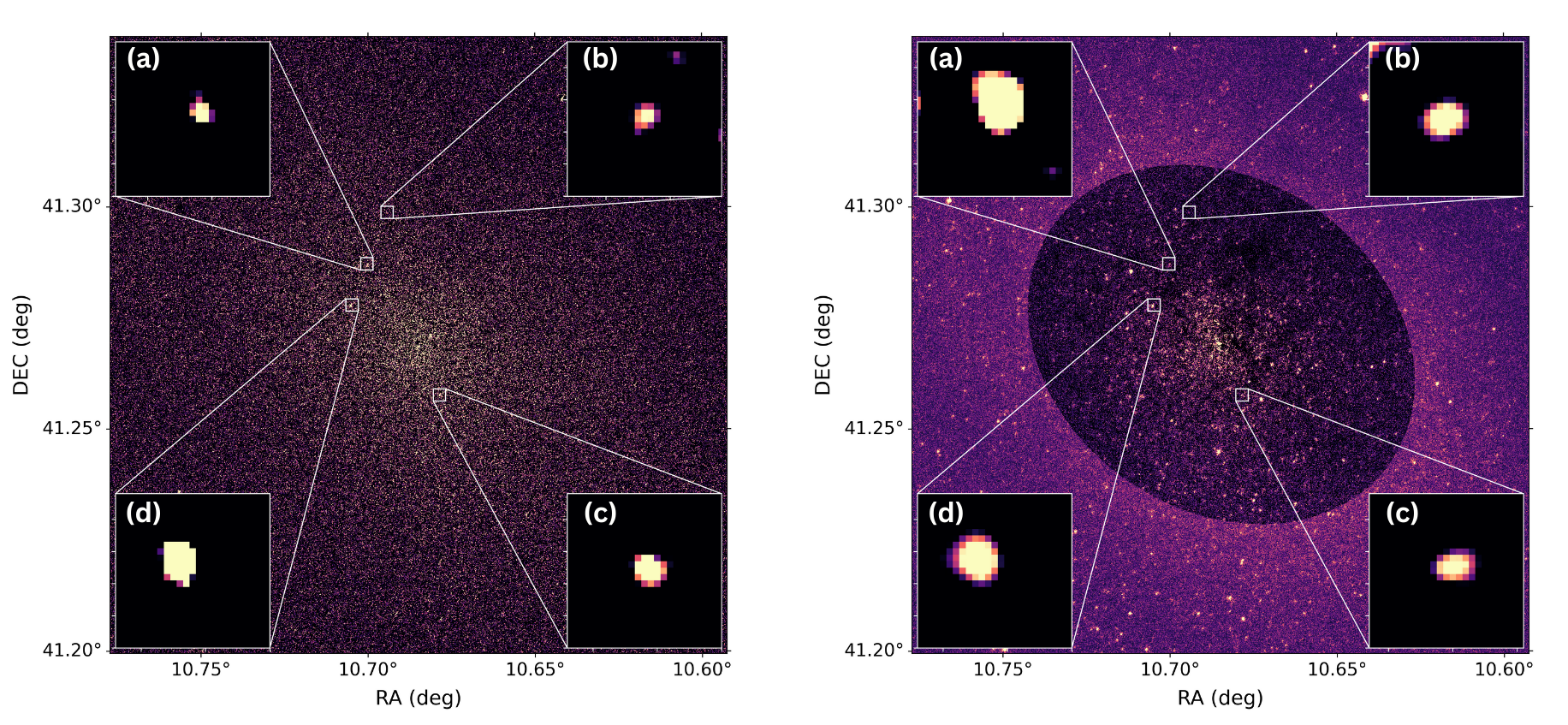}
    \caption{Difference images with inset showing detection of novae (a) M31N 2016-03e, (b) M31N 2015-05a, (c) M31N 2016-05b and (d) M31N 2016-03c. \textbf{Left}: Median-combined template subtraction method. \textbf{Right}: Isophote model subtraction method.}
    \label{fig:image_sub}
\end{figure*}

The F148W images of all the 19 M31 fields were combined together to produce a composite image of M31 in FUV. Individual images were normalized by their exposure times to bring them to the same scale. The normalized images were then tiled using \texttt{SWarp} \citep{Ber02_terapix}. The exposure array frames generated by \texttt{CCDLAB} were used as weights to take care of the edges and overlapping regions during tiling. The M31 FUV mosaic is shown in Figure~\ref{fig:M31_with_novae}.

\subsection{Removal of background contribution from the bulge}

The bulge of M31 is shrouded by galactic light. This light contamination must be removed to search for any point source in this region. Since traditional image subtraction techniques do not entirely remove the light excess in crowded fields, we have used two methods to create templates for detecting novae in the bulge region. 

The first method involves a median combination of multi-epoch images of the same field in the same filter. \citet{tandon20} showed that the sensitivity of the UVIT detectors has been constant over the years. Flux-scaling of the multi-epoch images thus only involved normalizing by their respective exposure times. Since UVIT is a space-based telescope, we expect the point spread function (PSF) to be constant across all epochs; hence, PSF matching was not performed before combining. The flux-scaled images were registered and then median-combined to make the template. This template was subtracted from the normalized image of each epoch to look for novae.

In the second method, we generated an isophote model of the bulge. We modeled the bulge of M31 by performing an isophote fit using the \texttt{isophote} module of the \texttt{photutils python} package \citep{Larry_2023}. The isophote model generated for a given range of semi-major axis (sma) ($\sim 400$ pixels = 2.78 arcmin) for each epoch was used as a template. Subtracting this template from its original image reveals all sources above the nuclear brightness level predicted by the model.

Both techniques were useful in detecting four novae, otherwise hidden in the bright M31 bulge. The difference images and the results from both methods are shown in Figure~\ref{fig:image_sub}. 

\subsection{Source detection and photometry}

A list of all M31 novae, as of 19 May 2022, is available at a website hosted by the Max Planck Institute for Extraterrestrial Physics \footnote{\href{https://www.mpe.mpg.de/~m31novae/opt/m31/M31_table.html}{https://www.mpe.mpg.de/$\sim$m31novae/opt/m31/M31\_table.html}} \citep{Pietsch_2010b}. We performed forced photometry at the locations of the novae listed to check for their detection in the UVIT data. We also looked into novae that erupted after the latest UVIT observations (Nov 2020), for which the data are publicly available, to check for pre-eruption UV magnitudes or upper limits. All the detected novae were carefully inspected visually to confirm the detections. Sources in crowded fields, where source confusion could occur within a 5~arcsec region of the nova, were rejected. We used a detection threshold of $3\sigma$ to eliminate faint sources. The limiting magnitude in the F148W filter for deep exposures ($>$ 10~ks) was found to be around 23~mags for a $3\sigma$ detection near the central region of M31. The detection limit for regions away from the center was around 0.5 mag deeper. The average limiting magnitude for each image is given in Table~\ref{obs_table}. However, it should be noted that the limiting magnitude varies spatially in an image. All the detected sources are shown in the UVIT F148W mosaic image in Figure~\ref{fig:M31_with_novae}, and their details are mentioned in Table~\ref{catalog}.

Availability of multiple images in the same field and the same filter encouraged us to sum them and look for fainter novae in the deeper exposures. Summing the images increased the limiting magnitude of the images by 0.2--0.4 mags (see Table~\ref{obs_table}). We found three old novae, namely 1985-09c, 1998-09e, and 2010-03a, in the co-added frame of the central region of M31. All three novae were fainter than 23 mag (see Table~\ref{catalog}). The FUV and NUV magnitudes of detected novae and upper limits of all undetected novae in the deepest available images in each filter were estimated, and have been provided in the appendix \ref{sec:appendix} Table~\ref{tab:master_table}.

\cite{leahy_2020} and \cite{Devaraj_2023} found that the PSF of UVIT images is better represented by Moffat functions rather than elliptical or circular Gaussian. However, neither Moffat nor Gaussian PSF could accurately estimate the flux from stars, which was underestimated due to the presence of extended PSF wings in the UVIT images \citep{tandon17b, leahy_2020}. In such cases, aperture photometry gives the best results, but due to crowded fields of M31, aperture photometry is not a feasible solution. PSF photometry was performed using an elliptical Moffat function to fit the sources. Standard routines in \texttt{IRAF} \citep{tod93} were used for photometric measurements. To accurately measure the flux, we performed aperture photometry with an aperture of 15 pixels ($\sim 4 \times $FWHM) and PSF photometry with a PSF radius of 11 pixels ($\sim 3 \times $FWHM) of isolated bright sources to determine the aperture correction term. This correction took care of the emission from the extended PSF wings. It was then applied to the magnitudes of the novae obtained by PSF photometry. The zero-point calibrations in the AB system were adopted from \citet{tandon20}. We have not corrected the magnitudes for extinction.

For the novae in the central fields, we tried aperture and PSF photometry on both sets of difference images. In aperture photometry, the background subtraction led to inconsistencies, whereas in PSF photometry, the PSF model fitting diverged. We thus performed PSF photometry with aperture correction of the bulge novae on the original science images using the same technique we used for the other novae. It ensured the measurements were consistent for all novae reported in this work. 

The photometry of all the detected sources in various filters at different epochs is given in Table~\ref{catalog}.

\begin{ThreePartTable}

\begin{TableNotes}
\item \textbf{Notes} \\
$\dag$ Also reported by \cite{Leahy_2021} \\
$^o$ Also detected around outburst (within $\sim$ 100 days after outburst and $\sim$ 10 days before outburst) \\
$^q$ Also detected at quiescence (more than 100 days after outburst or 30 days before outburst) \\
$^*$ $t_2$ determined from linear interpolation of the optical data around $(\rm m_{max}-2)$ \\
$\ddag$ Observation date for co-added images is the midpoint of the first and last observation \\
\item \textbf{References} - {\footnotesize  (1) \cite{ovcharov_2015}, (2) \cite{fabrika_2015}, (3) \cite{delvaux_2015}, (4) \cite{williams_2015}, (5) \cite{williams_2015a}, (6) \cite{darnley_2016}, (7) \cite{kasliwal_2018}, (8) \cite{hornoch_2016}, (9) \cite{hornoch_2016a}, (10) \cite{fabrika_2016}, (11) \cite{hornoch_2016b}, (12) \cite{fabrika_2016a}, (13) \cite{hornoch_2016c}, (14) \cite{valcheva_2016}, (15) \cite{fabrika_2016b}, (16) \cite{conseil_2018}, (17) \cite{campaner_2019}, (18) \cite{soraisam_2019}, (19) \cite{zhang_2020}, (20) \cite{williams_2020}, (21) \cite{hornoch_2020}, (22) \cite{williams_2020a}, (23) \cite{zhang_2020a}, (24) \cite{valcheva_2020}, (25) \cite{soraisam_2020}, (26) \cite{zhang_2020b}, (27) \cite{srivastav_2020}, (28) \cite{belligoli_2020}, (29) \cite{fabrika_2020}, (30) \cite{balcon_2020}, (31) \cite{hornoch_2020a}, (32) \cite{tomaney_1996}, (33) \cite{pietsch_2010}, (34) \cite{cao_2012}, (35) \cite{ovcharov_2013}, (36) \cite{hornoch_2013}, (37) \cite{kasliwal_2018a}, (38) \cite{darnley_2018}, (39) \cite{zhang_2018}, (40) \cite{valcheva_2018}, (41) \cite{hornoch_2018}, (42) \cite{williams_2019}, (43) \cite{zhang_2018a}, (44) \cite{hornoch_2019}, (45) \cite{hornoch_2019a}, (46) \cite{conseil_2019}, (47) \cite{williams_2019a}, (48) \cite{hornoch_2019b}, (49) \cite{zhang_2019}, (50) \cite{lee_2019}, (51) \cite{carey_2019}, (52) \cite{dahiwale_2020}, (53) \cite{soraisam_2019a}, (54) \cite{nordin_2019}, (55) \cite{hornoch_2019c}, (56) \cite{madrigal_2019}, (57) \cite{williams_2019b}, (58) \cite{fabrika_2019}, (59) \cite{zhang_2019a}, (60) \cite{jiang_2019}, (61) \cite{hornoch_2019d}, (62) \cite{williams_2020b}, (63) \cite{zhang_2020c}, (64) \cite{hornoch_2020b}, (65) \cite{zhang_2020d}, (66) \cite{fremling_2020}, (67) \cite{hornoch_2016d}, (68) \cite{martin_2018}, (69) \cite{hornoch_2015}, (70) \cite{Ovcharov_2015a}, (71) \cite{hornoch_2016e}. (72) \cite{Hornoch_2016f}, (73) \cite{Hornoch_2016g}, (74) \cite{Darnley_2016a}, (75) \cite{Clark_2024}, (76) \cite{Shumkov_2017}, (77) \cite{Ciardullo_1987}, (78) \cite{Sharov_2000}, (79) \cite{Pietsch_2010a}, (80) \cite{Hornoch_2010} 
\item For more references check \href{https://www.mpe.mpg.de/~m31novae/opt/m31/M31_table.html} {https://www.mpe.mpg.de/~m31novae/opt/m31/M31\_table.html}}

\end{TableNotes}

\begin{longtable*}[c]{lccccclll}

\caption{List of M31 novae detected in UVIT images and their F148W mags. Additional photometric data of these novae at different epochs, and in other filters will be available in the published version.}

\label{catalog} \\
\toprule
\toprule
Name & RA & DEC & Discovery  & Observation & Mag (F148W) & Class & $t_2$ & Ref. \\
M31N & (HH:MM:SS) & (DD:MM:SS) & Date (JD) & Date (JD) & AB & (Phase) & (days)& \\ 
\midrule
\endfirsthead

\multicolumn{9}{c}{Table \thetable~ continued from previous page} \\
\endhead

\endfoot
\insertTableNotes
\endlastfoot

\multicolumn{9}{c}{Detected at post outburst quiescence} \\
\midrule 
2015-02a $^\dag$ & 00:42:33.06 & +41:13:08.90 & 2457047.810 & 2457671.8566 & 23.27 $\pm$ 0.24 & Fe II & -- & 1-3 \\
2015-10b $^\dag$ & 00:43:15.34 & +41:20:16.60 & 2457321.460 & 2457671.8566 & 22.54 $\pm$ 0.13 & Fe II &  100.3 $\pm$ 75.8 (R) & 4,5,75 \\
2016-02b & 00:44:37.03 & +41:42:26.40 & 2457430.370 & 2457430.3700 & 22.01 $\pm$ 0.12 & Fe II & -- & 6 \\
2016-03b & 00:42:19.51 & +41:11:13.70 & 2457446.280 & 2457671.8567 & 22.16 $\pm$ 0.14 & Nova &  41.5 $\pm$ 7.1 (R) & 67,68,75 \\
1994-09b & 00:42:58.52 & +41:15:49.80 & 2449622.800 & 2457671.8566 & 21.88 $\pm$ 0.13 & -- & --  & 32 \\
2010-12b & 00:42:31.08 & +41:27:20.30 & 2455540.620 & 2457671.8566 & 21.64 $\pm$ 0.09 & -- & 3$(R)^*$  & 33,34 \\
2013-10c & 00:43:09.32 & +41:15:41.60 & 2456574.800 & 2457671.8566 & 24.32 $\pm$ 0.30 & Nova & 5.5 $\pm$ 1.7 (R) & 35,36,75 \\
2015-12d & 00:43:07.79 & +41:22:19.30 & 2457387.220 & 2457671.8566 & 21.72 $\pm$ 0.10 & -- & -- & -- \\
2018-12a & 00:42:40.65 & +41:11:08.00 & 2458455.320 & 2458788.0143 & 23.31 $\pm$ 0.14 & Fe II & -- & 39-42 \\
2018-12d & 00:43:24.62 & +41:20:22.30 & 2458474.090 & 2458788.0143 & 23.33 $\pm$ 0.12 & Nova & -- & 42,43 \\
2019-06b & 00:42:41.73 & +41:17:53.70 & 2458661.560 & 2458788.0143 & 22.75 $\pm$ 0.14 & -- & -- & 44 \\
2019-07c $^\dag$ & 00:43:00.90 & +41:19:19.40 & 2458672.510 & 2458788.0143 & 21.53 $\pm$ 0.06 & -- & -- & 45 \\
2019-08b & 00:44:18.25 & +41:32:47.20 & 2458725.440 & 2459167.9797 & 22.00 $\pm$ 0.08 & Fe II & $>$57($r'$), $>$53($g'$) $^*$  & 46,47 \\
2015-05a & 00:42:46.60 & +41:17:55.3 & 2457153.56 & 2457671.8566 & 23.66 $\pm$ 0.31 & Nova & -- & 69,70 \\
2016-03c & 00:42:49.14 & +41:16:40.1 & 2457430.24 & 2457671.8566 & 21.85 $\pm$ 0.13	& Nova & -- & 71,68 \\
2016-03e & 00:42:48.05 & +41:17:13.4 & 2457464.27 & 2457671.8566 & 22.78 $\pm$ 0.18 & -- & -- & 72 \\
2016-05b & 00:42:42.88 & +41:15:27.4 & 2457536.55 & 2457671.8566 & 22.89 $\pm$ 0.29 & Fe II & -- & 68,73,74\\
2017-06g & 00:46:17.94 & +41:56:26.1 & 2457922.69 &	2458061.9118 & 20.87 $\pm$ 0.12 & -- & -- & 76 \\
\midrule
\multicolumn{9}{c}{Detected at post outburst quiescence in coadded frames $\ddag$ } \\
\midrule
1985-09c & 00:42:44.50 & +41:17:35.0 & 2446320.74 &	2458550.6090 & 23.24 $\pm$ 0.10 & -- & -- & 77 \\
1998-09e & 00:42:49.53 & +41:16:47.9 & 2451083.28 &	2458550.6090 & 23.02 $\pm$ 0.09 & -- & -- & 78 \\
2010-03a & 00:42:47.74 & +41:17:01.4 & 2455257.25 & 2458550.6090 & 23.18 $\pm$ 0.10 & Nova & -- & 79,80 \\
\midrule
\multicolumn{9}{c}{Detected at pre outburst quiescence} \\
\midrule 
2019-12b & 00:42:20.47 & +41:15:00.00 & 2458833.090 & 2457671.8566 & 23.42 $\pm$ 0.25 & Fe II & 50($r'$), 27($g'$) $^*$  & 60-62 \\
2021-07d & 00:42:00.58 & +41:08:48.30 & 2459420.380 & 2458788.0143 & 22.99 $\pm$ 0.13 & -- & -- & -- \\
2019-11f $^o$ & 00:42:51.76 & +41:11:26.60 & 2458814.160 & 2457671.8566 & 22.95 $\pm$ 0.21 & -- & -- & 59 \\
2020-10b $^o$ & 00:42:30.67 & +41:21:28.50 & 2459127.230 & 2457671.8566 & 22.15 $\pm$ 0.13 & -- & -- & 63,64 \\
2020-10d $^o$ & 00:42:36.20 & +41:20:35.50 & 2459101.180 & 2457671.8566 & 23.12 $\pm$ 0.20 & -- & -- & 65 \\

\midrule
\multicolumn{9}{c}{Detected during outburst} \\
\midrule 
2016-08b $^\dag$ & 00:41:56.82 & +41:11:27.60 & 2457611.050 & 2457671.8566 & 20.71 $\pm$ 0.12 & -- & 42.6 $\pm$ 4.3 (R) & 7,8,75 \\
2016-08e & 00:43:53.29 & +41:26:21.60 & 2457630.420 & 2457704.1355 & 20.83 $\pm$ 0.06 & Fe II &  60.6 $\pm$ 16.7 (R) & 9,10,75 \\
2016-09a & 00:42:55.66 & +41:19:14.50 & 2457654.300 & 2457671.8566 & 23.56 $\pm$ 0.22 & Fe II &  57.0 $\pm$ 35.1 (R) & 11,12,75 \\
2016-09b & 00:42:17.29 & +41:09:44.40 & 2457657.660 & 2457671.8566 & 21.18 $\pm$ 0.09 & Fe II & 43.9 $\pm$ 17.3 (R) & 13-15,75 \\
2018-05a & 00:45:34.90 & +41:57:40.40 & 2458246.700 & 2458292.3095 & 20.84 $\pm$ 0.12 & -- & -- & 16 \\
2019-11a & 00:42:30.37 & +41:03:29.50 & 2458789.760 & 2458805.2565 & 22.92 $\pm$ 0.15 & Nova & 49($r'$) $^*$ & 17,18 \\
2020-09b & 00:42:46.30 & +41:13:35.50 & 2459117.200 & 2459174.7140 & 23.36 $\pm$ 0.18 & Fe IIb & -- & 19,20 \\
2020-09c & 00:43:08.38 & +41:08:50.00 & 2459121.790 & 2459174.7140 & 21.64 $\pm$ 0.08 & Fe II & 17($r'$), 15($g'$) $^*$  & 21,22 \\
2020-10f & 00:42:20.17 & +41:14:28.00 & 2459154.530 & 2459174.7140 & 20.74 $\pm$ 0.05 & Nova & -- & 23-25 \\
2020-11a & 00:42:46.79 & +41:07:03.10 & 2459155.080 & 2459174.7140 & 21.82 $\pm$ 0.09 & Nova & -- & 26,27 \\
2020-11c & 00:43:13.15 & +41:24:56.00 & 2459159.240 & 2459174.7140 & 19.05 $\pm$ 0.03 & Fe IIb & 20($r'$) $^*$ & 28-31 \\
2019-09b & 00:42:42.32 & +41:17:45.80 & 2458740.280 & 2458788.0143 & 21.72 $\pm$ 0.07 & He/N & -- & 48-50 \\
2019-10c $^\dag$ & 00:43:01.56 & +41:17:59.60 & 2458776.280 & 2458788.0143 & 21.09 $\pm$ 0.05 & Fe II & 18($r'$), $>$22($g'$) $^*$  & 54-58\\
2021-03a & 00:42:34.49 & +41:17:29.80 & 2459188.680 & 2459174.7140 & 22.89 $\pm$ 0.15 & -- & -- & 66 \\
2018-08b & 00:39:21.90 & +40:15:48.80 & 2458344.080 & 2458435.0755 & 20.92 $\pm$ 0.10 & Fe II & -- & 37,38 \\
2019-10a $^{q}$ & 00:42:23.53 & +41:11:52.20 & 2458759.300 & 2458788.0143 & 21.78 $\pm$ 0.08 & Nova & $>$18($r'$), $>$29($g'$) $^*$ & 51-53 \\

\bottomrule

\end{longtable*}
\end{ThreePartTable}

\section{Results and Discussion}

All the novae detected in the archival UVIT images are shown in Figure~\ref{fig:M31_with_novae} and listed in Table~\ref{catalog}. Nova outbursts can last for a wide range of duration, with a varied rate of decline ($t_2$ or $t_3$) of the light curve \citep{Warner_2008}. These parameters, in turn, depend on the physical nature of the system, i.e., the primary WD and the binary system. Theoretical models of \cite{Hachisu_2006} suggest the duration of the outburst depends on the WD mass and composition. \cite{Strope_2010} studied the optical light curves of around a hundred novae to look for various processes after the outburst capable of altering the light curve shapes. UV light curves have also been seen to vary in shape and duration. This is usually related to the super soft X-ray phase \citep{Page_2022}, which, in turn, is dependent on the WD mass \citep{Hachisu_2006, Hen11}. Due to a wide spectrum of speed class of novae, we decided to take 100 days as the cut-off for novae in outburst, corresponding to 7 mag decline from maxima for a 0.9 $M_{\odot}$ CO WD \citep{Hachisu_2006}. Additionally, considering pre-eruption dips and early UV flash, we consider novae detected within 20 days before the eruption to be in the outburst phase. For any other case, the detection is considered to be at quiescence. 

Overall, we have detected 42 novae in various fields of M31 observed by UVIT. Almost 90\% of the detected novae are concentrated within a 90~$\times$~90~arcmin region around the M31 center. This is expected because the nova rates follow the galactic light in optical/IR bands tracing the older stellar population of M31 \citep{Rector_2022}.  The following sections discuss the novae detected at different phases and their SEDs. 

\begin{figure}
    \centering
    \includegraphics[scale=0.8]{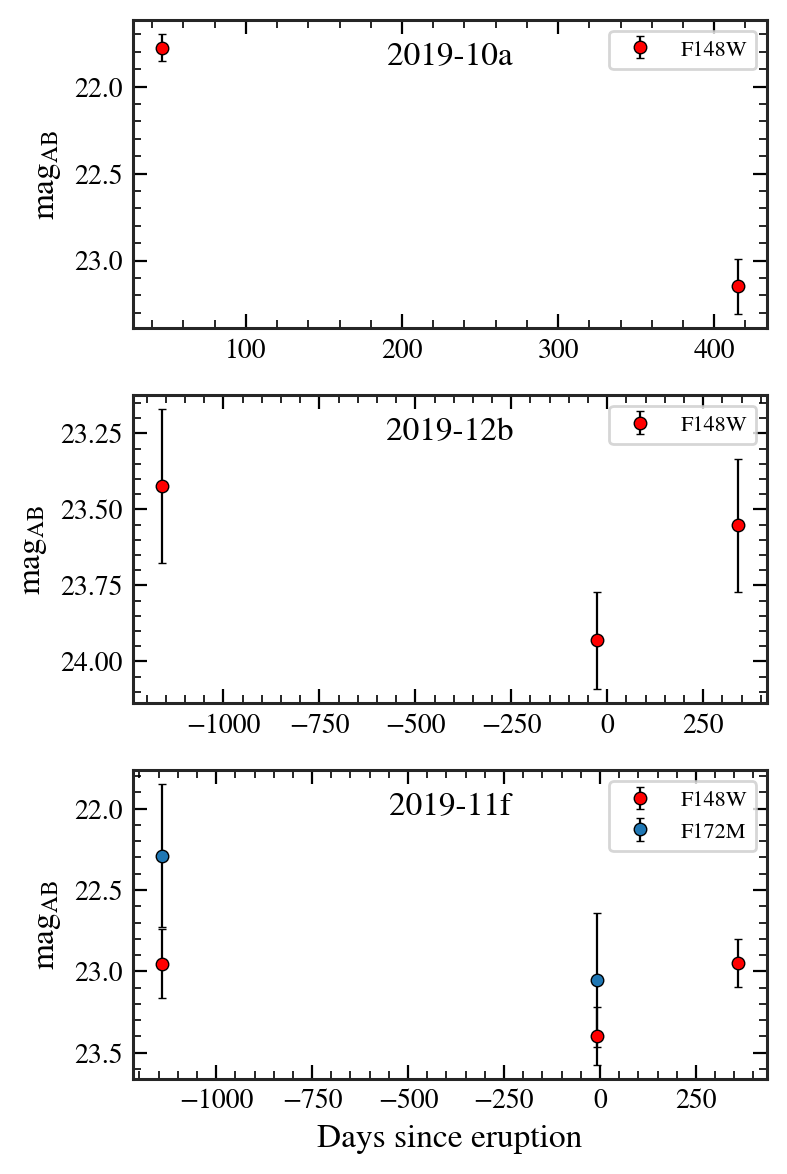}
    \caption{UVIT light curves of some interesting candidates}
    \label{fig:uvit_lc}
\end{figure}

\subsection{Novae detected at quiescence}
\label{sec:quiescent novae}
23 novae were detected during quiescence, 21 of which were detected post-outburst. 2021-07d was detected pre-outburst, and 2019-12b was detected both before and after the outburst. The quiescent F148W AB magnitudes of novae vary from 21.5 to 24 mags. Depending on the mass of the WD and accretion rate, the contribution in this filter could be due to both the WD and the disk. The quiescent magnitude also strongly depends on the system's orbital inclination \citep{Warner_2008}.

The light curve of nova 2019-12b (middle panel in Figure~\ref{fig:uvit_lc}) shows a variability in the F148W filter, notably a dip of $\sim$ 0.5 mag around 28 days before the eruption. \cite{Zamanov_23} pointed out that dips in light curves are caused by the accumulation of an optically thick, dense shell surrounding the WD prior to the nova eruption. Pre-eruption dips of around 1 mag have been noticed in other novae, such as T CrB in optical wavelengths \citep{ Schaefer_2023, Maslennikova_2024}. If caused by extinction due to an accumulated shell, these dips will be enhanced in the bluer bands. With the available data set, we cannot confirm the maximum drop in magnitude before the eruption. Nonetheless, a drop in FUV magnitude weeks before the eruption hints at the presence of a \textit{pre-eruption dip} in the light curve.

Most other novae detected more than once at post-outburst quiescence in the same filter do not show much variability between epochs. It strengthens the claim of nearly constant flux at quiescence by \cite{Selvelli_2013}. The disks of such systems have steady accretion rates and radiate at constant luminosity.

Among the old novae, only 2010-12b and 1994-09b have been detected at each epoch in multiple filters. Further, three more old novae, 2010-03a, 1998-09e, and 1985-09c, were detected in the co-added images. Detection of these old novae in multiple filters indicates a bright accretion disk. A high accretion rate during quiescence can make the disk brighter. The inclination of the system can also play a major factor in determining if such systems should be so bright long after the eruption. Another factor that could be responsible is the intrinsic brightness of the WD, which illuminates the accretion disk. The FUV flux in the F148W filter could even arise from novae systems with long SSS phases.

\subsection{Novae detected during outburst}
15 novae were detected within 100 days after the outburst, and 5 of them were detected within 20 days from the outburst. The only nova detected just before the outburst is 2021-03a, 14 days before its eruption. Most of the novae detected only during outbursts are bright ($\sim$ 19-20 mag) and fast. 6 novae were detected only once in the F148W filter.  4 novae were detected in multiple filters, and their SEDs are discussed in \S \ref{subsec:SED}. 

In the early outburst phase, the photospheric emission is the major source of UV photons. As the photosphere moves inward, the super-soft source dominates the FUV photon contribution. In the case of partially disrupted disks after eruption, contributions from the inner regions of the accretion disk to the FUV luminosity cannot be ruled out.

\subsection{Novae detected during quiescence and at outburst}
4 novae were detected both during quiescence and at outburst. 2019-10a was observed in the F148W filter at 46 and 415 days after the outburst, and it indicates a declining trend in its light curve as shown in the top panel of Figure~\ref{fig:uvit_lc}. 

2019-11f shows a pre-eruption dip similar to 2019-12b (see \S \ref{sec:quiescent novae}). In the F148W filter, it was observed 1140 and 8 days before the eruption and 360 days after the eruption. Additionally, the pre-eruption dip of 2019-11f is confirmed in the F172M filter observed 1140 and 8 days before the eruption. As shown in the bottom panel of Figure~\ref{fig:uvit_lc}, the pre-eruption dip is $\sim$ 8 days before the eruption. 

The other two novae, 2020-10b and 2020-10d, also show 0.4 and 0.7 mag variability, respectively, but the causes are unclear. However, we would like to point out that these variations are close to the limits of their respective errors.

\subsection{The spectral energy distributions (SEDs)}
\label{subsec:SED}
\begin{figure*}
    \centering
    \includegraphics[width=2\columnwidth]{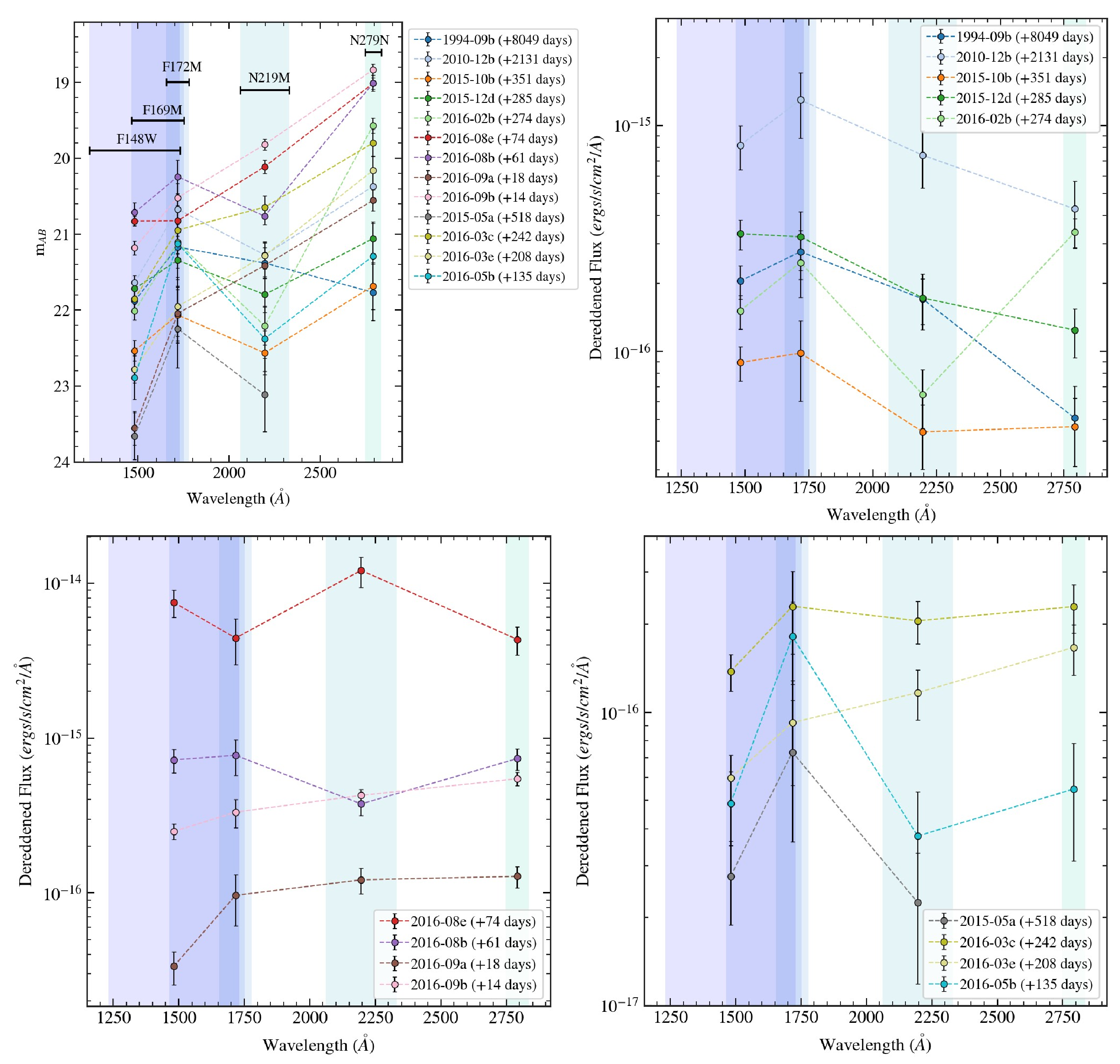}
    \caption{SED of novae detected after outburst are shown in the \textbf{top left} panel with AB magnitude in the y-axis and wavelength in the x-axis. The UVIT filter bands are also marked. The extinction-corrected flux of novae detected at post-outburst quiescence (\textbf{top right}), detected close to outburst (\textbf{bottom left}) and detected in the central region (\textbf{bottom right}) are also shown.}
    \label{fig:SED}
\end{figure*}

13 novae out of the 35 detected have near-simultaneous (within two days) observations in multiple filters in NUV and FUV. These data sets encouraged us to look into their SEDs. The observed SEDs are shown in the top-left subplot of Figure~\ref{fig:SED}. To correct for the extinction towards M31, we derived the $A_V$ values at each nova location from the M160 resolution dust mass maps of \cite{Draine_2014}, owing to its complete M31 coverage. However, compared to the PHAT survey of \cite{Dalcanton_2015}, the extinction values were overestimated by a factor of $\sim$ 2 (B.T.~Draine~2024, private communication). The $A_V$ values derived from \cite{Draine_2014} were corrected before applying to the SED of each nova. The UVIT AB magnitudes were converted to flux using the relations given in \cite{tandon17b}. We used the extinction functions from \cite{Cardelli_1989} to deredden the UV flux using the \texttt{extinction} \citep{Barbary_2021} package in \texttt{python}. The extinction-corrected SEDs in terms of $F_\lambda$ against $\lambda$ are shown in Figure~\ref{fig:SED}. 

The most interesting candidates are those detected at post-outburst quiescence, shown in the upper-right panel of Figure~\ref{fig:SED}. The power law dependence of $F_\lambda \propto \lambda^{- \alpha}$ was fitted to obtain the best-fit power-laws given in Table~\ref{tab:sed_slope}. The radiation from accretion disks follows a power-law with a slope of $\alpha=7/3$. The best-fit $\alpha$ values for the novae are within the error bars of this, indicating the presence of accretion disks \citep{frank_1992}. This confirms that most of the UV radiation originates from the accretion disk at quiescence \citep{Selvelli_2013}. Another point to note is that the flux in the F148W filter is comparatively less than in F172M. It could indicate that bluewards of 1600~\AA, the flux from the disk starts to diminish and could have a contribution from the WD's blackbody tail. 

\begin{table}[!ht]
    \caption{$\alpha$ values of novae with accretion disk signatures. $\alpha$ corresponds to the power-law between $F_\lambda$ vs $\lambda$ when errors in $A_V$ are not taken into account, whereas $\alpha_2$ corresponds for the same power-law when errors in $A_V$ are taken into account.}
    \label{tab:sed_slope}
    \begin{tabular}{ccc}
    \toprule
        M31N & $\alpha$ & $\alpha_2$ \\ 
        \midrule
        1994-09b & 2.71 $\pm$ 0.81 & 3.28 $\pm$ 0.79 \\ 
        2010-12b & 2.29 $\pm$ 0.002 & 2.29 $\pm$ 0.94 \\ 
        2015-10b & 2.01 $\pm$ 0.96 & 1.34 $\pm$ 1.32 \\ 
        2015-12d & 2.15 $\pm$ 0.33 & 1.92 $\pm$ 0.82 \\
    \bottomrule
    \end{tabular}
\end{table}

4 novae were caught during their eruption in multiple filters, and their SEDs are shown in the lower-left panel of Figure~\ref{fig:SED}. Two novae, 2016-09a and 2016-09b, were observed within 18 days of eruption. Their speed class suggests that they belong to a moderately fast category, and both were also confirmed to be in their Fe II phase (see Table~\ref{catalog}). Their SEDs indicate an increasing flux from FUV to NUV, hinting the peak to be in the optical region. The peak brightness of the nova eruption corresponds to the maximum extent of the photosphere. During the maxima, the peak of the SED moves to longer wavelengths, which could be in UV for very fast novae but also in optical or even IR for slow novae. As the photosphere recedes, hotter regions closer to the WD are revealed, and the SED peak moves blue wards. A SED with its peak in optical bands could indicate that these two novae were observed at a stage when their pseudo-photosphere was shrinking. 

The SEDs of the novae 2016-02b, 2016-08e, and 2016-08b deviate from the power law expected from accretion disks. One possible cause for this could be the presence of spectral features in the UV bands. The quiescent novae sample of \cite{Selvelli_2013} consisted majorly of C {\sc iv} 1550 \AA, along with some fainter features like N {\sc v} 1240 \AA, Si {\sc iv} 1400 \AA, He {\sc ii} 1640 \AA, O {\sc iii]} 1666 \AA, and N {\sc iii]} 1750 \AA~in emission. FUV absorption features arising from metals such as Si {\sc ii} 1190 \AA, Si {\sc ii} 1260 \AA, O {\sc i}+Si {\sc ii} 1303 \AA, C {\sc ii} 1335 \AA, Si {\sc iv} 1400 \AA, C {\sc iv} 1550 \AA, and N {\sc v} 1240 \AA~can also distort the SEDs. During outburst, however, the UV spectral features are mostly emission until the Fe curtains stage sets in. The list of emission lines and how they evolve have been described in \cite{Shore_2008}.

Another possibility for the deviation of the SED could be the improper determination of the extinction parameter, which would amplify any errors in the FUV bands as well as in the N219M band (since it covers the UV extinction bump at 2175 \AA). The SED of the quiescent nova 2016-02b could be under-corrected for extinction. The SED of 2016-08e might be over-corrected, whereas that of 2016-08b might have been under-corrected, both of which were captured during outburst. The over-correction in some of the SEDs of novae could imply that the sources are away from the galaxy's disk towards us, i.e., these sources suffer less extinction than predicted by the dust maps. Similarly, sources with under-correction may be located on the far side of M31, thus suffering more extinction than obtained from the dust maps.

The bulge of M31 in FUV is complicated. \cite{Leahy_2023} could fit the bulge with a single component but showed that an 8-component model fits best. Such a complicated bulge region will lead to contamination, dependent on the spatial location of the novae and on the different UV wavebands. Furthermore, the location of these novae inside the nuclear region is unknown, leading to uncertainties in determining their column densities and extinction. It thus becomes difficult to extract much information from the SEDs of novae in the central region of M31.

\section{Summary and Conclusion}
Astronomers have extensively studied novae in M31, in the optical for more than a century, and in X-rays over the last couple of decades to understand their distribution, rates, and evolution at different phases. However, UV surveys remain limited. This study utilizes multi-band AstroSat UV data to explore cataclysmic variables, novae in particular, during outbursts and quiescence. Below, we present the key findings from our investigation 
\begin{itemize}
    \item  We have examined 91 sets of UVIT images of M31 from 2016 onwards, uncovering at least one nova in 19 of those images. Over 80\% of the identified novae were situated within or close to the central bulge.
    \item We employed two image subtraction methods to eliminate the light from the galactic bulge, leading to the detection of four novae previously obscured by the brightness of the central region of M31. 
    \item A total of 42 novae were identified, with 15 observed across multiple filters in both FUV and NUV. Additionally, several novae were detected in multiple epochs. Magnitude upper limits were estimated for more than 1000 novae undetected in individual and co-added frames.
    \item During quiescence, the spectral energy distributions (SEDs) of novae exhibit indications of the accretion disk's influence, confirming its dominance in UV radiation during this phase. Analysis of the SEDs of two novae during outburst indicates they were observed while their photospheres were receding. 
    \item The FUV light curves of two novae exhibit a pre-eruption dip that can possibly be attributed to the accumulation of accreted material in a shell before the nova eruption. Multi-epoch photometry of the novae at quiescence indicates a near-constant magnitude, a sign of a steady accretion rate.  
\end{itemize}

Upcoming UV missions like INSIST \citep{Annapurni_2022}, CASTOR \citep{Patrick_2012}, UVEX \citep{Kulkarni_2021}, and ULTRASAT \citep{Shvartzvald_2024} offer the potential for detecting novae at different stages of their UV evolution. These observations can provide insights into disk activity, accretion processes, and the characteristics of WD during dormancy phases. Additionally, by employing high-frequency observations, we anticipate elucidating the UV flash preceding eruptions, as predicted by the theoretical models of \citet{kato_2022}, along with the rapid UV evolution post-outburst. We have initiated a UV survey of M31's central region using the FUV F148W filter on UVIT/AstroSat. Initial observations have revealed the detection of the newly identified RN M31N 2013-10c during our monitoring activities \citep{Basu_2024}. The FUV survey was designed to complement the simultaneous monitoring of M31 by GIT \citep{Harsh_2022} in $g',r'$ bands, which has detected numerous novae and even discovered the nova AT 2023tkw (GIT230919aa, \citealt{ravi_2023}). The integration of UV monitoring with multi-wavelength surveys such as ZTF \citep{Bellm_2019} and GIT in optical, XRISM in X-rays \citep{xrism_2020}, and the forthcoming Roman Space Telescope in IR \citep{Akeson_2019}, alongside Daksha in X-rays \citep{Bhalerao_2024}, promises to contribute significantly to resolving outstanding questions concerning nova phenomena.

\section{Acknowledgement}
We appreciate the anonymous reviewer's insightful comments and suggestions, which have enhanced the quality of this article. This work uses the archival UVIT data from the AstroSat mission of the Indian Space Research Organisation (ISRO). We thank the PIs of the proposals that were observed. The proposal IDs are the first seven characters of the observation IDs given in Table~\ref{obs_table}. We thank the UVIT payload operation centers for verifying and releasing the raw and processed data via the ISSDC data archive and providing the necessary software tools. GCA thanks the Indian National Science Academy for support under the INSA Senior Scientist Programme.  KS thanks the Indian Institute of Astrophysics for supporting her work while at IIA.

\facilities{ AstroSat (UVIT) }

\software{\texttt{CCDLAB} \citep{postma17}, 
            \texttt{IRAF v2.16.1} \citep{tod93}, 
            \texttt{Python v3.6.6} \citep{python09},  
            \texttt{NumPy} \citep{NumPy20}, 
            \texttt{SciPy} \citep{SciPy20}, 
            \texttt{Pandas} \citep{mckinney10}, 
            \texttt{Matplotlib} \citep{Hunter07}, 
            \texttt{photutils} \citep{Larry_2023},
            \texttt{SWarp} \citep{Ber02_terapix},
            \texttt{SAOImageDS9} \citep{ds9_2003}
            }

\appendix
\section{Catalog Table}
\label{sec:appendix}
\begin{table*}
\begin{rotatetable*}
\caption{UVIT catalog of all M31 novae detection and non-detection in each filter, along with available archival data. A sample table with ten rows is shown below. The full version of the table will be made available in machine-readable format in the published paper.}
\label{tab:master_table}
\hspace*{-22em}    
\resizebox{1.63\linewidth}{!}{%
    \begin{tabular}{ccccccccccccccc}
     \toprule
     Name & RA & DEC & UVIT field & Individual Detection & Discovery Date & Max mag & Filter & $t_2$ & Alternative names & F148W mag & F172M mag & F169M mag & N219M mag & N279N mag \\
     M31N & hh:mm:ss & dd:mm:ss & &     *            & JD             &         &        & days  &                   & AB        &   AB       &     AB    &    AB     &   AB     \\
    \midrule
    2022-04d & 00:38:48.05 & +40:15:26.7 & 15 & N & 2459695.52 & 16.5 & w & -- & AT2022iro & $>$ 23.34 &    & $>$ 24.30 & & \\
    2022-04c & 00:42:57.33 & +41:16:18.4 & 1 & N & 2459692.41 & 17.3 & w & -- & AT2022icc & $>$ 23.31 & $>$ 22.29 & $>$ 22.67 & $>$ 21.86 & $>$ 21.05 \\
    2021-07d & 00:42:00.58 & +41:08:48.3 & 1 & Y & 2459420.38 & 19.3 & w & -- & XM57MZ & 23.21 $\pm$ 0.09 & 23.76 $\pm$ 0.32 & 22.40 $\pm$ 0.17 & $>$ 22.08 & $>$ 21.53 \\
    2021-07c & 00:42:17.92 & +41:00:18.5 & 3 & N & 2459399.89 & 18.3 & r & -- & AT2021scd, ZTF21abjiotr & $>$ 22.83 & $>$ 22.07 & & $>$ 22.36 & $>$ 21.87 \\
    2020-10f & 00:42:20.17 & +41:14:28.0 & 1 & Y & 2459154.53 & 16.3 & R & -- & AT2020yky, XM42MZ & 21.99 $\pm$ 0.042 & $>$ 23.12 & $>$ 22.63 & $>$ 22.08 & $>$ 21.48 \\
    2020-10e & 00:42:37.35 & +41:20:52.2 & 1 & N & 2459144.24 & 16.3 & R & 4 & AT2020xyv, XM41MZ & $>$ 24.02 & $>$ 23.18 & $>$ 22.63 & $>$ 22.08 & $>$ 21.53 \\
    2020-01b & 00:42:52.34 & +41:16:13.1 & 1 & N & 2458877.04 & 17.40 $\pm$ 0.15 & R & 11.4 $\pm$ 2.5 & AT2020ber, XM24MZ & $>$ 22.96 & $>$ 22.37 & $>$ 22.60 & $>$ 21.61 & $>$ 20.68 \\
    2017-06g & 00:46:17.94 & +41:56:26.1 & 8 & Y & 2457922.69 & 18.1 & w & -- & MASTER OT J004617.94+415626.1 & 21.89 $\pm$ 0.10 & $>$ 23.28 &  & $>$ 22.06 & 20.70 $\pm$ 0.18 \\
    2016-11b & 00:42:42.81 & +41:12:33.4 & 1 & N & 2457720.26 & 16.03 $\pm$ 0.05 & R & 12.6 $\pm$ 1.8 & PNV J00424281+4112334 & $>$ 23.53 & $>$ 22.91 & $>$ 22.59 & $>$ 22.08 & $>$ 21.53 \\
    2016-03e & 00:42:48.05 & +41:17:13.4 & 1 & Y & 2457464.27 & 16.9 & w & -- & PNV J00424805+4117134 & 23.98 $\pm$ 0.21 &	23.59 $\pm$ 0.52 & $>$ 22.15 & 21.28 $\pm$ 0.18 & 20.16 $\pm$ 0.18 \\
    \vdots & \vdots & \vdots & \vdots & \vdots & \vdots & \vdots & \vdots & \vdots & \vdots & \vdots & \vdots & \vdots & \vdots & \vdots \\
    \vdots & \vdots & \vdots & \vdots & \vdots & \vdots & \vdots & \vdots & \vdots & \vdots & \vdots & \vdots & \vdots & \vdots & \vdots \\
    \bottomrule
    \end{tabular}
    }
    \textbf{Notes:} \\
    * Y indicates detection in at least one individual frame. N indicates non-detection in individual frames. \\
    The UVIT magnitudes were estimated from the co-added images where more than one epoch of data was available in the same field and filter. In other cases, magnitudes were estimated from the individual frames. \\
    Blank fields in the UVIT magnitude columns indicate the absence of data of that field in that filter. \\
    
\end{rotatetable*}    
\end{table*}

\newpage

\bibliography{mybib}{}
\bibliographystyle{aasjournal}

\end{document}